\documentclass[conference, 9pt]{IEEEtran}
\usepackage{spconfa4,graphicx}
\usepackage{graphicx}
\usepackage{multirow}
\usepackage{amsmath,amssymb,amsfonts}
\usepackage{amsthm}
\usepackage{mathrsfs}
\usepackage[figuresright]{rotating}
\usepackage{xcolor}
\usepackage{textcomp}
\usepackage{manyfoot}
\usepackage{booktabs}
\usepackage{algorithm}
\usepackage{algorithmicx}
\usepackage{algpseudocode}
\usepackage{program}
\usepackage{listings}
\usepackage{cite}

\begin{document}

\title{Online Segmented Beamforming via Dynamic Programming}

\name{Manan Mittal $^1$, Ryan M. Corey $^{2}$, Diego Cuji$^1$, John R. Buck $^3$, Andrew C. Singer $^1$ \thanks{Funded in part by US Navy, Office of Naval Research under award N00014-23-1-2133}}
\address{Stony Brook University$^1$, University of Illinois Chicago$^2$, University of Massachusetts Darthmouth$^3$}

\maketitle

\begin{abstract}
In dynamic acoustic environments characterized by time-varying interferers and moving sources, effective beamforming requires accurately identifying stationary regions over time. Traditional Capon beamformers rely on the instantaneous ensemble covariance matrix, which is inaccessible in practice. Practical implementations overcome this by estimating the sample covariance matrix (SCM) through averaging over a block of temporal samples. However, in non-stationary settings, a naive batch approach fails. Moving interferers smear the SCM, causing the beamformer to place nulls in outdated locations while failing to track newly active interferers, thereby degrading its nulling capabilities. To address this fundamental limitation, an Online Segmented Beamformer is proposed. This algorithm incorporates data-driven temporal segmentation to causally minimize output power while dynamically adapting the SCM estimation windows to local stationarity. By framing the problem through the lens of dynamic programming, the proposed method tracks abrupt environmental changes and resets covariance estimates in real-time. We validate the performance of this framework in a complex, reverberant simulated acoustic environment and in highly reverberant real world experiments, demonstrating its superiority over fixed-window adaptive methods.
\end{abstract}

\section{Introduction}

Beamforming algorithms are fundamental tools in array signal processing, acting as spatial filters that process measurements from a distributed sensor array. Their primary objective is to enhance target signals arriving from a specific direction while simultaneously suppressing background noise and spatially localized interference. Beamforming is the basis for many modern systems from speech enhancement in smart devices to robust target tracking in underwater sonar and radar systems. The standard benchmark for adaptive spatial filtering is Capon’s Minimum Variance Distortionless Response (MVDR) beamformer \cite{capon}. The MVDR formulation seeks a weight vector that minimizes the total output power of the array, subject to a linear constraint that maintains a distortionless, unity response in the look direction of the desired signal. By strictly constraining the target response, the algorithm ensures that any minimization of the total output power is achieved exclusively through the suppression of noise and interfering signals. Theoretically, this mathematically elegant solution relies on the true ensemble covariance matrix (ECM) of the environment \cite{capon}. 

However, the practical utility of the Capon beamformer is severely constrained by the reality that the ECM is never known a priori and it must be estimated from finite, noisy data. This estimation introduces a fundamental bias-variance tradeoff governed almost entirely by the length of the integration window. To minimize estimation variance and ensure the sample covariance matrix (SCM) is well-conditioned, a long integration window is desirable. Conversely, real-world acoustic environments are profoundly non-stationary. Interferers appear and vanish, sources physically move, and multipath reverberation patterns evolve as the scene changes. In such dynamic settings, relying on a long integration window inevitably spans multiple distinct acoustic states. The resulting SCM represents a smeared temporal average that does not correspond to any singular, physical state of the environment. Consequently, a beamformer derived from this average wastes the array's spatial degrees of freedom, placing nulls where interferers used to be while remaining blind to current interference. Traditional approaches to non-stationary beamforming typically attempt to address this issue through fixed-memory mechanisms, such as sliding windows, exponential forgetting factors, or multi-rate adaptive processing techniques \cite{cox_multirate}. While these allow for some degree of adaptation, they are inextricably tied to hyperparameters that implicitly assume a specific, constant timescale of stationarity. 

To overcome the inherent limitations of fixed-memory estimation, this work introduces the Online Segmented Beamformer. We propose that the optimal integration time should not be a static hyperparameter, but rather a dynamic variable inferred directly from the statistical structure of the data itself. Inspired by the Online Segmented Recursive Least Squares (OSRLS) algorithm \cite{osrls}, our causal method operates sequentially, evaluating at each time step whether to extend the current stationary segment (thereby improving variance) or to declare a change point and reset the covariance estimate (thereby preventing bias from an outdated state). By explicitly modeling the piecewise-stationary nature of dynamic acoustic scenes, the proposed beamformer effectively acts as a universal estimator capable of tracking abrupt environmental transitions with high fidelity.

\section{Signal Model}

Consider an array comprising $N_m$ sensors arbitrarily distributed in an acoustic environment containing $N_s$ localized sources. We assume a centralized processing architecture where all sensor measurements are jointly available for synchronous processing.

Let $s_j[t]$ denote the source signal emitted by the $j^{\text{th}}$ acoustic source. The signal received at sensor $m$ is modeled as the convolution of $s_j$ with the room or ocean acoustic impulse response $h_{m,j}[t]$, combined with additive, spatially uncorrelated background noise $v_m[t]$. The observed signal $x_m[t]$ at time $t$ is formalized as:
\begin{equation}
x_m[t] = \sum_{j=1}^{N_s} (h_{m,j} \ast s_j)[t] + v_m[t],
\end{equation}
where $\ast$ denotes discrete-time convolution.

The broadband beamformer applies a linear finite impulse response (FIR) filter of length $L$ to each sensor channel and sums the resulting outputs. To facilitate analysis, this operation is expressed in vector notation. The most recent $L$ samples from all sensors are concatenated into a single spatio-temporal observation vector $\mathbf{x}[t] \in \mathbb{R}^{N_m L}$:
\begin{equation}
\mathbf{x}[t] = \begin{bmatrix} x_1[t], \dots, x_1[t{-}L{+}1], \dots, x_{N_m}[t], \dots, x_{N_m}[t{-}L{+}1] \end{bmatrix}^\top.
\end{equation}
The corresponding beamformer coefficients are stacked into a time-varying weight vector $\mathbf{w}[t] \in \mathbb{R}^{N_m L}$. The scalar beamformed output $z[t]$ at time $t$ is thus given by the inner product:
\begin{equation}
z[t] = \mathbf{w}[t]^\top \mathbf{x}[t].
\end{equation}

The instantaneous second-order statistics of the received data are captured by the correlation matrix $\mathbf{R}[t] = \mathbb{E}[\mathbf{x}[t] \mathbf{x}[t]^\top] \in \mathbb{R}^{N_m L \times N_m L}$, where $\mathbb{E}[\cdot]$ denotes the expectation operator. 

Finally, to extract the target signal without distortion, the beamformer utilizes a constraint vector, denoted by $\boldsymbol{\nu}$. In anechoic or plane-wave environments, this is the standard analytic steering vector. In complex reverberant spaces, this constraint vector represents the Relative Transfer Function (RTF), characterizing the relative gains and delays of the desired source across the array \cite{consolidated}. 

\section{Adaptive to Segmented Beamforming}

\subsection{The Capon Ideal and Practical Limitations}
The fundamental insight of the Capon beamformer is the minimization of output power to suppress interference. The theoretically optimal weight vector $\mathbf{w}_{\text{opt}}$ is the solution to the following constrained optimization problem:
\begin{equation}
    \min_{\mathbf{w}} \quad J(\mathbf{w}) = \mathbb{E}\big[ \mathbf{w}^\top \mathbf{R}[t] \mathbf{w} \big] \quad \text{subject to} \quad \mathbf{w}^\top \boldsymbol{\nu} = 1.
\end{equation}
Solving this yields the classic MVDR solution:
\begin{equation}
    \mathbf{w}_{\text{opt}}[t] = \frac{\mathbf{R}[t]^{-1}\boldsymbol{\nu}}{\boldsymbol{\nu}^\top \mathbf{R}[t]^{-1}\boldsymbol{\nu}}.
\end{equation}

In practical real-time implementations, the ensemble covariance $\mathbf{R}[t]$ is unavailable and must be approximated by the sample covariance matrix (SCM), $\mathbf{S}[t]$, computed over a window of $K$ prior samples:
\begin{equation}
    \mathbf{S}[t] = \frac{1}{K} \sum_{n=t-K}^{t-1} \mathbf{x}[n]\mathbf{x}[n]^\top.
\end{equation}
The adaptive MVDR weights are then computed via $\mathbf{w}[t] = \frac{\mathbf{S}[t]^{-1}\boldsymbol{\nu}}{\boldsymbol{\nu}^\top \mathbf{S}[t]^{-1}\boldsymbol{\nu}}$. 

When a standard implementation computes this SCM over an observation window, it effectively minimizes the sum of squared outputs over that specific batch of data. In dynamic environments, using a static window length $K$ forces a compromise. If $K$ is too large, the beamformer fails to track moving sources. If $K$ is too small, estimation noise dominates, leading to signal distortion.

\subsection{Proposed Method: The Segmented Beamformer Approach}
We posit that the standard formulation is a restrictive case of a more general optimization problem: the temporally segmented minimum variance beamformer. 

Instead of constraining the filter to a single weight vector over a fixed window, we allow the observation timeline to be partitioned into discrete segments. Let $\mathcal{E}(i, j)$ denote the minimum output power achievable by a single, constant Capon weight vector over an interval from sample $i$ to sample $j$. If one were to minimize total output power without any constraints, the trivial solution would create a new segment for every single sample, driving the error to zero but entirely eliminating the covariance averaging necessary for true interference suppression. 

To prevent this degenerate overfitting, a regularization parameter $C$ is introduced to penalize model complexity. The objective becomes finding a sequence of partitions that minimizes the total piecewise output power plus the cost of each new segment. 

\subsection{Proposed Online Causal Beamformer}
While a global segmentation can be found retrospectively via dynamic programming, it requires $O(T^2)$ operations and is inherently non-causal. To enable sequential, real-time processing, we adopt an online approximation analogous to the Online Segmented Recursive Least Squares (OSRLS) method. 

The Online Segmented Beamformer maintains the most recently detected segmentation point, denoted $t_p$, and assumes that prior boundary is correct. At each incoming sample, the algorithm compares the cost of extending the current segment (maintaining the current covariance matrix) against the cost of initiating a new segment at some intermediate point. The recursive cost function evaluated at time $t$ is expressed as:
\begin{equation}
    E(t) = \min_{t_p \le i \le t} \Big( \mathcal{E}(i, t) + C + \mathcal{E}(t_p, i-1) \Big) + E(t_p-1).
\end{equation}

Operationally, the algorithm manages a set of candidate MVDR beamformers running in parallel. Each candidate hypothesizes a different start time for the current stationary segment. The inverse covariance matrix $\mathbf{S}^{-1}$ and corresponding weight vector $\mathbf{w}$ for each candidate are efficiently updated sample-by-sample using the Woodbury matrix identity. 

Whenever the accumulated tracking cost of a candidate starting at a later time, including the transition penalty $C$, drops below the cost of the currently active segment, the algorithm declares a structural change point. The beamformer immediately "switches" to the new optimal candidate, effectively flushing the outdated covariance matrix and rapidly adapting to the new acoustic regime. This greedy approximation ensures the algorithm operates efficiently in linear or constant time (if a maximum search window is imposed) while remaining highly responsive to scene dynamics. We provide an algorithmic description of the proposed online segmented beamformer in the algorithm \ref{alg:OSRLS_BF}

\begin{algorithm}[htbp]
\caption{Online Segmented Beamformer}\label{alg:OSRLS_BF}
\begin{algorithmic}[1]
\Require $\mathbf{X} \in \mathbb{R}^{p \times T}$ (samples), $\boldsymbol{\nu} \in \mathbb{R}^{p}$ (Steering), $C$ (Penalty), $\delta$ (Loading), $\tau$ (Min segment length)
\Ensure $\mathbf{z} \in \mathbb{R}^{T}$ (Output), $\mathcal{I}$ (Partition indices)

\State \textbf{Initialize per-start states:}
\For{$i = 0$ \textbf{to} $T{-}1$}
    \State $\mathbf{S}^{-1}[i] \gets \mathbf{I}_p / \delta$
    \State $\mathbf{u}[i] \gets \boldsymbol{\nu} / \delta$ \Comment{Numerator state: $\mathbf{S}^{-1}\boldsymbol{\nu}$}
    \State $\rho[i] \gets \boldsymbol{\nu}^\top \mathbf{u}[i]$ \Comment{Denominator state: $\boldsymbol{\nu}^\top \mathbf{S}^{-1}\boldsymbol{\nu}$}
    \State $\mathbf{w}[i] \gets \mathbf{u}[i] / \rho[i]$
    \State $J[i] \gets 0$ \Comment{Accumulated segment cost (output power)}
\EndFor
\State $\mathcal{I} \gets [0]$, $\text{cur} \gets 0$
\State $E[-1] \gets 0$, $\mathbf{z} \gets \mathbf{0}_T$

\For{$n = 0$ \textbf{to} $T{-}1$}
  \State $\mathbf{x} \gets \mathbf{X}[:, n]$
  \State $\mathbf{z}[n] \gets \mathbf{w}[\text{cur}]^\top \mathbf{x}$ \Comment{Filter output using active model}

  \State $E_{\min} \gets \infty$, $\text{best} \gets \text{cur}$

  \For{$i = \text{cur}$ \textbf{to} $n$} \Comment{Update candidates}
    \State $\mathbf{k} \gets \mathbf{S}^{-1}[i] \mathbf{x} \,/\, (1 + \mathbf{x}^\top \mathbf{S}^{-1}[i] \mathbf{x})$ \Comment{Woodbury update}
    \State $\mathbf{U} \gets \mathbf{k} \mathbf{x}^\top \mathbf{S}^{-1}[i]$
    \State $\mathbf{S}^{-1}[i] \gets \mathbf{S}^{-1}[i] - \mathbf{U}$
    
    \State $\mathbf{u}[i] \gets \mathbf{u}[i] - \mathbf{U} \boldsymbol{\nu}$ \Comment{Update MVDR weights}
    \State $\rho[i] \gets \boldsymbol{\nu}^\top \mathbf{u}[i]$
    \State $\mathbf{w}[i] \gets \mathbf{u}[i] / \rho[i]$
    
    \State $y \gets \mathbf{w}[i]^\top \mathbf{x}$ \Comment{Update cost}
    \State $J[i] \gets J[i] + y^2$

    \State $E_{\text{total}} \gets E[i{-}1] + C + J[i]$
    \If{$E_{\text{total}} < E_{\min}$}
      \State $E_{\min} \gets E_{\text{total}}$
      \State $\text{best} \gets i$
    \EndIf
  \EndFor

  \State $E[n] \gets E_{\min}$

  \If{$(\text{best} - \text{cur}) > \tau$}
    \State $\text{cur} \gets \text{best}$ \Comment{Switch to new optimal segment}
    \State Append $\text{best}$ to $\mathcal{I}$
  \EndIf
\EndFor

\State \Return $\mathbf{z}, \mathcal{I}$
\end{algorithmic}
\end{algorithm}

\section{Simulations}
\begin{figure}[t]
\centering
\includegraphics[width=\linewidth]{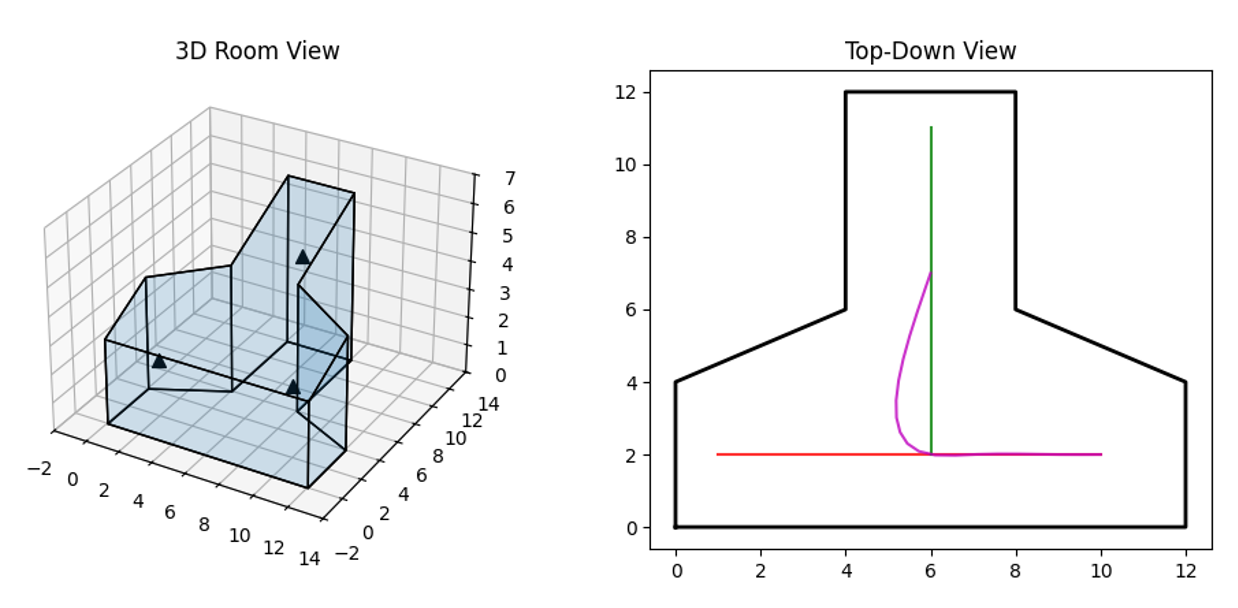} 
\caption{The simulated room acoustics environment. On the left a 3-d depiction of the simulated room. On the right, a top-down view of the three moving source trajectories.}
\label{fig:room_sim}
\end{figure}

To evaluate the algorithm's efficacy in complex, reverberant environments, we conducted simulations using a custom acoustic simulator built upon the pyroomacoustics framework \cite{pra}, extended to support continuous source trajectories. The simulated environment consists of a complex polygon room with a reverberation time ($T_{60}$) of $200$ ms. The sensing infrastructure comprises three circular microphone arrays, each containing 10 sensors. The spatial layout of the room and the array geometry are depicted in Fig. \ref{fig:room_sim}. Within this space, we define a static target source located at coordinates $[6.0 m, 2.0m, 1.5m]$ and three dynamic interferers. These interferers follow complex, non-linear trajectories through the room, creating a continuously shifting spatial covariance structure. 

Processing is executed in the frequency domain via a Short-Time Fourier Transform (STFT) with a frame size of $1024$ samples and an overlap of $512$. The switching beamformer operates independently across each frequency bin. To handle the reverberant nature of the environment, we use a Relative Transfer Function (RTF) steering vector rather than a free-field model. The RTF for the desired source is estimated for each bin using the covariance whitening method \cite{cw_rtf, 8937151}. 

We benchmark the online segmented framework against a range of fixed-length sliding-window MPDR beamformers, with window lengths ranging from highly reactive (20 frames) to highly stable (1200 frames). As summarized in Table  \ref{tab:results}, the switching beamformer achieves the highest performance across both the objective metrics. Specifically, it provides superior Scale-Invariant Signal-to-Distortion Ratio (SI-SDR), indicating more effective null-steering against the moving interferers. Furthermore, the Perceptual Evaluation of Speech Quality (PESQ) and Short-Time Objective Intelligibility (STOI) scores demonstrate that the switching mechanism minimizes the audible artifacts and musical noise often introduced by aggressive, short-window adaptive filters. This confirms that the algorithm successfully balances the rapid tracking required for moving sources with the statistical smoothing necessary for high-fidelity audio reconstruction.

\begin{table}[ht]
\centering
\caption{Performance Comparison in room acoustics simulation}
\label{tab:results}
\begin{tabular}{lccc}
\toprule
\textbf{Method} & \textbf{PESQ} & \textbf{SI-SDR Gain (dB)} \\ 
\midrule
\textbf{OSRLS (Proposed)} & \textbf{1.08} & \textbf{5.91} \\
MPDR (Win=20)   & 1.06 & 3.10 \\
MPDR (Win=70)   & 1.06 & 2.88 \\
MPDR (Win=120)  & 1.06 & 2.70 \\
MPDR (Win=200)  & 1.06 & 2.64 \\
MPDR (Win=400)  & 1.06 & 2.63 \\
MPDR (Win=1200) & 1.06 & 2.57 \\
\bottomrule
\end{tabular}
\end{table}

\section{Experiments}
To evaluate the proposed Online Segmented Beamformer in a physical, highly reverberant environment, we utilized the Massive Distributed Microphone Array Dataset \cite{corey_massive_2019}. This dataset contains acoustic measurements and speech recordings captured by 160 microphones distributed throughout a large conference room measuring 13 m by 9 m, with a reverberation time ($T_{60}$) of approximately $800$ ms. 

The sensing infrastructure is highly heterogeneous, comprising two types of arrays: wearable microphone arrays containing 16 sensors each (placed on mannequins) and tabletop arrays containing 8 sensors each (housed in voice-assistant-style enclosures). For our experiments, we constructed a spatially distributed subarray by randomly selecting 40 channels uniformly from the 160 available microphone positions. This random subset mimics an ad-hoc distributed acoustic sensor network, presenting a challenging scenario for traditional beamforming due to the lack of uniform geometry and extreme multi-path effects.

The target and interference signals consist of 60-second continuous speech clips derived from the VCTK corpus. To simulate a realistic, dynamic cocktail party scenario, the dataset provides recordings of these speech signals played simultaneously through multiple loudspeakers. Furthermore, ambient background noise recorded in the conference room with no active loudspeakers was added to the mixtures to ensure realistic signal-to-noise ratios.

For this evaluation, we restricted the acoustic scene to four active talkers (Talkers 1 through 4) distributed spatially across the room. The beamforming task requires extracting the desired talker while continuously adapting to suppress the remaining three interfering talkers and the ambient room noise. The Relative Transfer Functions (RTFs) were estimated using the isolated exponential frequency sweep (chirp) recordings provided in the dataset for each loudspeaker location.

The proposed OSRLS algorithm was applied to the a randomly selected 40-channel mixture in the STFT domain using a frame size of 1024 samples and 50\% overlap. We compared its performance against the standard MPDR beamformer utilizing fixed sliding windows of varying lengths.

\begin{figure}[htbp]
\centering
\includegraphics[width=\linewidth, height = 15cm]{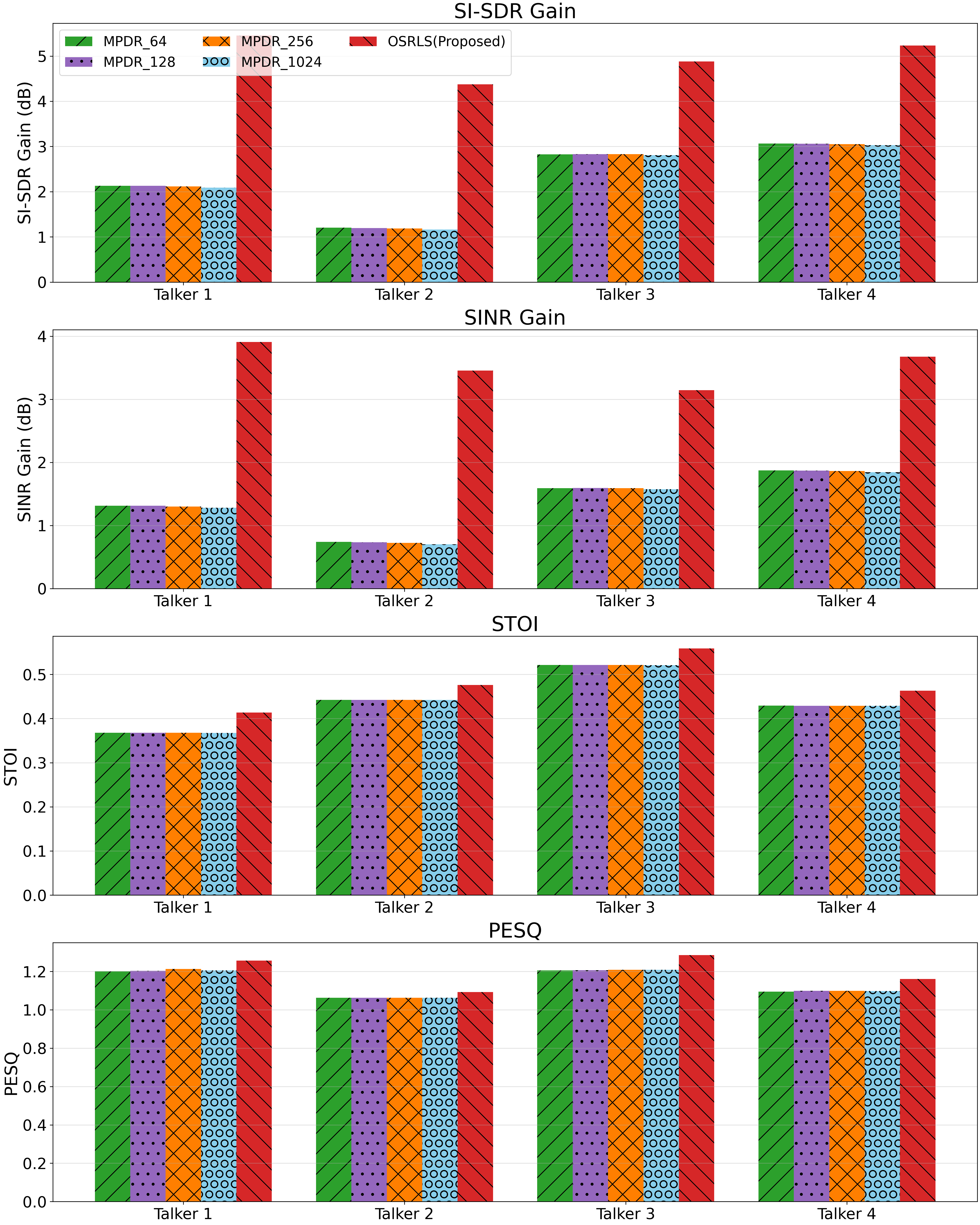} 
\caption{Performance comparison of the proposed OSRLS beamformer versus standard MPDR across different target speakers.}
\label{fig:performance_plot}
\end{figure}

As shown in Figure \ref{fig:performance_plot}, the real-world highly reverberant data corroborates the findings from our simulations. The fixed-window MPDR formulations struggle to balance the bias-variance tradeoff; short windows suffer from severe covariance estimation noise due to the lack of spatial averaging, while long windows fail to track the subtle phase shifts and dynamic multi-path reflections inherent in a real room. The Online Segmented Beamformer autonomously partitions the observation timeline, resulting in superior SI-SDR and PESQ metrics. By flushing outdated interference information, the proposed method maintains deep spatial nulls against the competing talkers without distorting the target speech.

% \begin{table}[ht!]
% \centering
% \caption{Performance Comparison on Real-World Data (40-Channel Distributed Array)}
% \label{tab:real_results}
% \begin{tabular}{lccc}
% \toprule
% \textbf{Talker} & \textbf{Method} & \textbf{PESQ} & \textbf{SI-SDR Gain (dB)} \\ 
% \midrule
% \multirow{5}{*}{T01} 
%  & \textbf{OSRLS (Proposed)} & \textbf{1.26} & \textbf{5.45} \\
%  & MPDR (Win=64)   & 1.20 & 2.13 \\
%  & MPDR (Win=128)  & 1.20 & 2.13 \\
%  & MPDR (Win=256)  & 1.21 & 2.12 \\
%  & MPDR (Win=1024) & 1.21 & 2.09 \\
% \midrule
% \multirow{5}{*}{T02} 
%  & \textbf{OSRLS (Proposed)} & \textbf{1.09} & \textbf{4.38} \\
%  & MPDR (Win=64)   & 1.06 & 1.20 \\
%  & MPDR (Win=128)  & 1.06 & 1.19 \\
%  & MPDR (Win=256)  & 1.06 & 1.19 \\
%  & MPDR (Win=1024) & 1.06 & 1.16 \\
% \midrule
% \multirow{5}{*}{T03} 
%  & \textbf{OSRLS (Proposed)} & \textbf{1.29} & \textbf{4.88} \\
%  & MPDR (Win=64)   & 1.21 & 2.83 \\
%  & MPDR (Win=128)  & 1.21 & 2.83 \\
%  & MPDR (Win=256)  & 1.21 & 2.83 \\
%  & MPDR (Win=1024) & 1.21 & 2.81 \\
% \midrule
% \multirow{5}{*}{T04} 
%  & \textbf{OSRLS (Proposed)} & \textbf{1.16} & \textbf{5.23} \\
%  & MPDR (Win=64)   & 1.09 & 3.07 \\
%  & MPDR (Win=128)  & 1.10 & 3.06 \\
%  & MPDR (Win=256)  & 1.10 & 3.05 \\
%  & MPDR (Win=1024) & 1.10 & 3.03 \\
% \bottomrule
% \end{tabular}
% \end{table}

\section{Conclusion}
In this work, a principled framework for adaptive beamforming in non-stationary acoustic environments has been presented. By identifying the fundamental limitation of standard adaptive beamformers—namely the reliance on fixed integration windows that smear distinct acoustic states—the Online Segmented Beamformer was developed. This algorithm functions as a universal estimator, dynamically adapting its effective memory length to the underlying rate of environmental change without requiring hyperparameter tuning tied to an assumed stationarity rate. By leveraging dynamic programming principles to causally evaluate and reset covariance estimates, the beamformer maintains tracking agility while preserving necessary statistical averaging. Validations in complex, reverberant scenarios demonstrated the robustness of the proposed method, providing a parameter-free alternative to traditional adaptive beamforming exceptionally well-suited for time-varying acoustic applications.

\bibliographystyle{IEEEtran}
\bibliography{ieee-bibliography}

\end{document}